\documentclass[letterpaper, 11 pt, peerreviewca,onecolumn]{ieeeconf}

\usepackage{courier}
\usepackage{hyperref}
\usepackage{graphicx,subfigure}
\usepackage[scriptsize]{caption}
\usepackage{epstopdf}
\usepackage{tikz}
\usepackage{multirow}
\usepackage{float}
\usepackage{amsmath,amssymb,amsfonts}
\usepackage{graphicx}
\usepackage{flushend}
\usepackage[ruled]{algorithm}
\usepackage{algpseudocode}
\usepackage{bm}
\usepackage{algorithm,float}
\usepackage{algpseudocode}
\usepackage{verbatim}
\newtheorem{theorem}{Theorem}
\newtheorem{lemma}{Lemma}

\newtheorem{definition}{Definition}
\newtheorem{remark}{Remark}
\newtheorem{example}{Example}
\makeatletter

\newcommand{\Rmnum}[1]{\expandafter\@slowromancap\romannumeral #1@}

\makeatother

\title{\LARGE \bf  
Koopman-based Deep Learning for Nonlinear System Estimation
}
\author{Zexin Sun, Mingyu Chen, John Baillieul}
\makeatother

\begin{document}
	\maketitle
	\thispagestyle{empty}
	\pagestyle{empty}
	
	\let\thefootnote\relax\footnotetext{\noindent\underbar{\hspace{0.8in}}\\
		Zexin Sun is with the Division of Systems Engineering at Boston University. Mingyu Chen is with the Department of Electrical and Computer Engineering at Boston University. John Baillieul is with the Departments of Mechanical Engineering, Electrical and Computer Engineering, and the Division of Systems Engineering at Boston University, Boston, MA 02215. The authors may be reached at {\tt \{zxsun, mingyuc, johnb\}@bu.edu}. }
	
	\begin{abstract}
Nonlinear differential equations are encountered as models of fluid flow, spiking neurons, and many other systems of interest in the real world.  Common features of these systems are that their behaviors are difficult to describe exactly and invariably unmodeled dynamics present challenges in making precise predictions. In this paper, we present a novel data-driven linear estimator based on Koopman operator theory to extract meaningful finite-dimensional representations of complex nonlinear systems. The Koopman model is used together with deep reinforcement networks that learn the optimal stepwise actions to predict future states of nonlinear systems. Our estimator is also adaptive to a diffeomorphic transformation of the estimated nonlinear system, which enables it to compute optimal state estimates without re-learning.
		
	\end{abstract}
	\section{Introduction}
Deriving analytical models to accurately characterize the intricate dynamics of real-world physical and biological systems represents a longstanding scientific challenge. From celestial motions \cite{moser2001stable} to neuronal spike patterns \cite{ostojic2011spiking}, the complexity of natural phenomena often precludes precise mathematical specification---a problem compounded by measurement imprecision and stochasticity. Extracting useful information from nonlinear models such as those arising in advanced applications of control theory requires innovative approaches to state estimation and prediction, \cite{simon2006optimal}. 

While particular processes have been encoded as nonlinear differential equations such as Navier-Stokes to describe the fluid flow or the Van der Pol equation model in seismology, these formulations exhibit complex dynamics that do not admit explicit solutions \cite{tu2013dynamic}.  Alternative data-driven approaches, such as \cite{schmid2010dynamic,williams2015data} must confront the inherent difficulties of distilling governing laws solely from previous observations. A promising methodological paradigm sidesteps direct system identification through the framework of Koopman operator theory. By lifting nonlinear dynamics into a high-dimensional space, simpler linear representations can enable more robust forecasting, analysis, and control. \cite{brunton2019notes}. However, meaningful finite representations must still be identified — a tradeoff between interpretability, accuracy, and dynamical relevance. Recent research \cite{yeung2019learning} focused on spectral analysis principally employs dictionary-based learning of Koopman operator eigenfunctions to approximate nonlinear system dynamics. While all linear embeddings learned by neural networks are usually efficiently characterized, their training process is still time-consuming. However, this idea motivates us to combine machine learning methods with Koopman operators. Other related work, like \cite{bevanda2022learning,bevanda2022diffeomorphically} adopts the strong assumption that the nonlinear system is globally exponentially stable and it has a diffeomorphic linear representation. We address such challenges with a new approach to synthesizing Koopman frameworks based on reinforcement learning. We first construct representations by extracting dominant terms from polynomial approximation that may render imperfectly due to neglected nonlinearities. By incorporating closure terms adapted via reinforcement learning, the resulting estimator compensates for unmodeled dynamics. Similar work has been reported in \cite{mowlavi2023reinforcement} which investigates reduced-order modeling of nonlinear systems through dynamic mode decomposition (DMD) \cite{tu2013dynamic}, compensating for simplified dynamics via neural networks. By sacrificing precise system characterization, their approach achieves dimension reduction relative to high-fidelity numerical discretizations. However, reliance on data for suitable performance renders the training process critical.

In this context, we propose novel contributions on three fronts. First, we develop a linear state estimator based on  extended dynamic mode decomposition (EDMD) \cite{williams2015data}. By performing singular value decomposition of a data-based approximation of the Koopman operator, we obtain a reduced order model in which meaningful features are preserved. Second, a supplemental compensation term to account for residual inaccuracies in the EDMD is introduced.  This correction term is learned by means of an approach to deep reinforcement learning based on a deep deterministic policy gradient (DDPG) algorithm. Optimal policies then map current system states and their estimation to actions. Finally, we consider transfer learning for generalization beyond training scenarios. The learned policies are shown to retain nearly optimal error bounds across a range of newly introduced dynamics.


 \section{Preliminaries}
 \label{section_math}
	Consider an autonomous nonlinear system
\begin{equation}
		\label{nonlinear_basic}
		\bm{\dot x}={ \bm{f}(\bm{x})},\\
\end{equation}
where $ \bm{f}(\cdot):\mathbb{R}^n\rightarrow\mathbb{R}^n$ is assumed to be Lipschitz continuous but whose exact form is unknown. In this context, all vectors and functions are denoted by bold symbols.
The flow of this dynamical system can be expressed as $\bm{F}^{t}(\bm{x_0}):=\bm{x_0}+\int_{0}^{{t}}\bm{f}(\bm{x}(\tau))d\tau $ that we assume is a unique solution on $[0,\infty)$ from the initial point $\bm{x_0}$. Then, the semigroup Koopman operator $\mathcal{K}^{ t}$ associated with the flow map $\bm{F}^{t}$ is defined by $\mathcal{K}^{ t}_{\bm{f}}T_{\bm{x}}=T_{\bm{x}}\circ \bm{F}^t$, where $	\mathcal{K}^t_{\bm{f}}$ is a linear operator and $T_{\bm{x}}$ is called an observation function. The symbol $\circ$ denotes function composition. An approach to constructing a class of such functions will be given in the next section.
We can formally represent the evolution of our Koopman dynamical system by writing $\frac{d}{dt} T_{\bm x}=\mathcal{K}_{\bm{f}}T_{\bm{x}}$, where $\mathcal{K}_{\bm{f}}$ is the  infinitesimal generator of the one-parameter family of $	\mathcal{K}^t_{\bm{f}}$. A principal incentive to employ the Koopman operator framework is its capacity to simplify the analysis of dynamical systems via eigen-decomposition of the associated linear operator, which enables an efficient characterization of the evolution of observables chosen to represent the nonlinear system.  
\begin{definition}
	A Koopman eigenfuction (KEF) is an observable $\phi$ that satisfies $\mathcal{K}^t_{\bm{f}}\phi=e^{\lambda t}\phi$, where $e^\lambda$ is called the Koopman eigenvalue (KE). 
	\end{definition}
 
Given such an eigen-decomposition, the observation functions may be rendered as $T_{\bm{x}}=\sum_{j=1}^{\infty}v_j\phi_j $, where $v_j$ is called the Koopman mode (KM). 
However, in practice, the KEF, KF, KM triplets are difficult to identify because we have no access of the exact function $\bm{f}$. A data-driven method for approximating the tuples of the Koopman operator called {\em extended dynamic mode decomposition} (EDMD) is proposed. EDMD captures nonlinear dynamics in a linear space, thereby providing more than merely local information about system behavior while retaining the linear characteristics of matrix decompositions. It is an extension of dynamic mode decomposition (DMD), which has been used to approximate the KEs and KMs by collecting a series of measurements of the system and performing a singular value decomposition \cite{brunton2015closed}. Suppose the measurements obtained from $(\ref{nonlinear_basic})$ are sampled at the end of each time interval $\Delta t$ and a discrete time dynamic system is associated with $\bm{f}$ for a fixed time interval $\Delta t$ whose time- evolution is
\begin{equation}
\label{eq:discrete}
\bm{x_{k+1}}=\bm{F}(x_k).
\end{equation}
Then, at the sample times, $\mathcal{K}_{\bm{f}}=\mathcal{K}_{\bm{F}}$ and the KEs of $\bm{F}$ are $\lambda^F_i=e^{\lambda_i\Delta t}$. Therefore, 
$
 T_{\bm{x}}(\bm{x}_{k+1})=\mathcal{K}_{\bm{F}}T_{\bm{x}}(\bm{x}_{k})=\sum_{j=1}^{\infty}v_j\lambda^F_j\phi_j.
$

 Consider the nonlinear system (\ref{nonlinear_basic}) and its discretized form (\ref{eq:discrete}) with sampling interval $\Delta t$, under the assumption that the function $\bm{f}$ is nonlinear, time-invariant, and unknown. Based on the above Koopman operator theoretic framework, the problem is formulated to identify a linear embedding that captures the dynamics of (\ref{nonlinear_basic}). In what follows, the objective entails synthesizing an estimator architecture that leverages the derived linear embedding together with a function approximator such as a neural network that learns from the sequences of observations to predict future states of the nonlinear system. 

\section{Deep Reinforcement Learning-based Estimator}
In this section, we construct an estimator to learn from sequences of post-reference trajectories of a discrete nonlinear system (\ref{eq:discrete}). The techniques include two phases: the first one constructs an EDMD-based linear estimator of the nonlinear system via singular value decomposition to extract dominant observable functions; the second phase adds a term to the linear estimator based on reinforcement learning that compensates for the discrepancies.

\subsection{EDMD-based Linear Approximation}
The EDMD algorithm is a data-driven technique that requires two input datasets \cite{williams2015data}.  
Suppose we have $M$ equal-length sequences of state trajectories $\{\bm{\tau_1}, \cdots, \bm{\tau_M}\}$, where $\bm{\tau_i} = \{\bm{x_{i, 0}}, \cdots, \bm{x_{i, L}}\}$ represents the states of trajectory $i$, which satisfies $\bm{x_{i, l+1}} = \bm{F(x_{i, l})}$ for $l \in  \{0,\cdots, L-1\}$.
To process the EDMD, we first construct a dataset of snapshot pairs $\mathcal{D} = \{( \bm{x^j}, \bm{y^j})\}_{j=1}^{M\times L}$, where $\bm{x^{j}}$ belongs to any state in the trajectory and $\bm{y^j}$ represents the next state of $\bm{x^{j}}$ in the same trajectory.
Then, we select polynomial products as a dictionary of observable functions \[\bm{\Psi}=\{\psi_1,\psi_2,\cdots,\psi_{N}\},N\gg n\] to approximate the KEFs. From Stone-Weierstrass types of results we know that polynomials are good (dense) approximators of continuous functions. Here gives a simple example. 
\begin{example}
    Assume that the system has the form
\begin{equation}
\begin{split}
\dot{\bm{x}}=\begin{bmatrix}
  -0.7 x_1 \\
  -0.3(x_2-x_1^2)
\end{bmatrix}.
\end{split}
    \label{ex1}
\end{equation}
It has a finite dimensional Koopman operator by applying the polynomial-based dictionary $\bm{\Psi(x)}=\{x_1,x_2,x_1^2\}$. Then, (\ref{ex1}) can be
rewritten as an equivalent linear system 
\[\dot{\bm{\Psi}}(\bm{x})=\begin{bmatrix}
  -0.7&  0& 0\\
 0 & -0.3 &0.3 \\
 0 & 0 &-1.4
\end{bmatrix}\bm{\Psi(x)}.\]
\end{example}

Next, we define a vector-valued function $\bm{\Psi^T(x)}=[\psi_1(\bm{x}),\psi_2(\bm{x}),\cdots,\psi_{N}(\bm{x})]^T$, where the superscript T denotes the matrix transpose. It is assumed that the vector formed by the first $n$ elements in $\bm{\Psi}$ is \[[{\psi_1}(\bm{x});{\psi_2}(\bm{x});\cdots;\psi_n(\bm{x})]=\bm{x}\]
Given these prerequisites, the EDMD algorithm \cite{brunton2015closed} can be applied to the system (\ref{eq:discrete}), which finds a set of \emph{full-state observables}  whose dynamics evolves linearly. Then, the observables defined by snapshot pairs $(\bm{x^j},\bm{y^j})$ in $\mathcal{D}$ are arranged into two data matrices, $\bm{\Psi(X)}$ and $\bm{\Psi(Y)}$:
\begin{equation*}
\begin{split}
\bm{\Psi(X)}&=\begin{bmatrix}
 \mid & \mid &  &\mid\\
  \bm{\Psi(x^1)}& \bm{\Psi(x^2)} & \cdots &\bm{\Psi(x^{M\times L})}\\
 \mid&\mid  &  &\mid
\end{bmatrix},\\
\bm{\Psi(Y)}&=\begin{bmatrix}
 \mid & \mid &  &\mid\\
  \bm{\Psi(y^1)}& \bm{\Psi(y^2)} & \cdots &\bm{\Psi(y^{M\times L})}\\
 \mid&\mid  &  &\mid
\end{bmatrix}.
\end{split}
\end{equation*}
The best fit linear operator is given by 
\begin{equation}
\label{eq:A}
A=\arg\min_{A}||A\bm{\Psi(X)}-\bm{\Psi(Y)}||^2=\bm{\Psi(Y)\Psi(X)^\dagger},
\end{equation}
where $\dagger$ denotes Moore-Penrose inverse of a matrix. 
Given the EDMD approximation, the nonlinear system can be written via the Kooperman Operator $\mathcal{K}_{\bm f}$ and be approximated by a linear model
\begin{equation}
 \begin{split}
        \bm{\Psi^*(\bm{x_{k+1}})}&=\mathcal{K}_{\bm f}\bm{\Psi^*(\bm{x_k})},\\
	\bm{\Psi(\bm{x_{k+1}})}&\approx A\bm{\Psi(\bm{x_k})},\label{eq:linear}
 \end{split}
\end{equation}
where $\bm{\Psi^*}$ denotes the ideal oberservable function that is used to construct all KEFs. Meanwhile, in \cite{klus2015numerical,korda2018convergence} conditions have been given whereby the sample-based approximation $\bm{\Psi}$ approaches $\bm{\Psi^*}$ and the true Koopman operator $\mathcal{K}_{\bm f}$ is approached in the limit as $N\rightarrow\infty$ and $M\rightarrow\infty$.


However, in practice, observables with high-degree polynomials produce a high-dimensional state space in which it is intractable to represent and spectrally decompose the Koopman Operator. 
To tackle this problem, instead of computing $A$ by (\ref{eq:A}) we perform singular value decomposition of $\bm{\Psi(X)}$ and extract its leading singular vectors to form a projection space for $A$, along the lines of the steps in DMD algorithm \cite{tu2013dynamic}, i.e., $\bm{\Psi(X)}\approx U\Sigma V^T$, where $U,V\in \mathbb{R}^{N\times r}$ are semi-unitary and $\Sigma\in\mathbb{R}^{r\times r}$. Then, $A=\bm{\Psi(Y)}V\Sigma^{-1}U^T$ and $A_r=U^TAU=U^T\bm{\Psi(Y)}V\Sigma^{-1}$ is the reduced-order matrix, whose leading $r$ eigenvalues and eigenvectors are the same as $A$. Therefore, we can write 
\begin{equation*}
\begin{split}
    \bm{\tilde \Psi(x_{k+1})}&\approx A_r\bm{\tilde \Psi(x_{k})}\\
    \bm{\Psi(x_{k+1})} &= U\bm{\tilde \Psi(x_{k+1})},
    \end{split}
\end{equation*}
where $\bm{\tilde \Psi}$ is in the reduced-order state space.

By designing the reduced-order estimator based on $\bm{\tilde \Psi(x)}$, it potentially leads to significant approximation errors when relying solely on such linear EDMD embedding. Furthermore, state measurements are often corrupted by noise, further degrading the accuracy of estimators derived from such linear approximations. Therefore, our parametric estimator according to (\ref{eq:linear}) is corrected to be 
\begin{equation}
	\label{eq:e_linear}
	\begin{split}
 \bm{\tilde\Psi(\bm{\hat x_{k+1}})}&=A_r\bm{\tilde\Psi(\bm{\hat x_{k})}}+\bm{a_k}\\
	\end{split}
\end{equation}
where $\bm{a_k}:\mathbb{R}^{n}\to \mathbb{R}^{N}$ is a function that can predict the difference between the linear embedding and nonlinear system. Since the first $n$ elements in $\bm{\Psi}$ are given by $\bm{\Psi}_{1:n}(\bm{x})=\bm{x}$ to represent the full state observation and reconstruction, given a $\bm{\tilde\Psi(\hat x_{k+1})}$, we may predict $\bm{\hat{x}_{k+1}}=P\bm{\Psi({\hat x_{k+1}}})=PU\bm{\tilde\Psi({\hat x_{k+1}}})$, where $P=[\bm{I}_n,\mathbf{0}_{N-n}]$ and $\bm{I}_n$ is the n-dimensional identity matrix. Given that the dictionary is pre-designed and the reconstruction $\bm{x}$ only requires extracting the first $n$ entries from the dictionary, this process can be executed efficiently. The approach significantly reduces the computational time compared to directly fitting a model using a neural network, a benefit that has been corroborated by simulation results. 

The remaining challenge is how to describe the nonlinear behavior $\bm{a_k}$. 
It suffices to note that $\bm{a_k} = \bm{\tilde\Psi(F(\bm{ x_k}))}- A_r\bm{\tilde\Psi(\bm{\hat x_k)}}$ for all $\bm{x_k},\bm{\hat x_k}\in \mathcal{X}$. However, since the function $\bm{f}$ is unknown and the state space $\mathcal{X}$ is continuous, the precise value of function $\bm{F}$ is unattainable. Meanwhile, the measurement of $\bm{x_k}$ may have noise at each time step. A more feasible approach is to learn \(\bm{\hat{a}}\) based on the measurement data $\bm{\tilde x_k}$ of (\ref{eq:discrete}) that has the form 
\begin{equation*}
    \bm{\tilde x_k}=\bm{x_k}+\bm{\omega_k},
\end{equation*}
where $\bm{\omega_k}$ is the measurement noise, which is often assumed to be Gaussian noise with zero mean. 
In this context, we use the deep reinforcement learning on the estimator and thereby learn the residual error $\bm{a_k}$ capable of any given state.

    \subsection{Deep Reinforcement Learning-based Estimator}
  
  Reinforcement learning (RL), which involves continuous interaction between an agent and its environment, can be utilized to obtain such a correction. The agent receives the current state ($\bm{s_k}$) from the environment and subsequently selects an action ($\bm{a_k}$) from a large, continuous action space based on a nonlinear policy $\pi_\theta$. The environment transitions into a new state ($\bm{s_{k+1}}$) and returns a reward to the RL agent that reflects the quality of the chosen action. Through this agent-environment interaction, the reinforcement learning agent can learn to optimize the policy for maximum cumulative reward.

  However, traditional Q-learning methods cannot be used in this estimation problem because of the continuous action space and the number of trajectories of the system being limited.  We therefore adopt the deep deterministic policy gradient (DDPG) algorithm \cite{lillicrap2015continuous}, which in turn adopts the actor-critic structure and uses neural networks to fit nonlinear functions. The Q network (critic) and policy network (actor) are represented by $Q$ and $\mu$ with weights $\theta^{Q}$ and $\theta^{\mu}$ accordingly. Similar to the DQN algorithm we have been used in \cite{sun2023emulation}, DDPG also has target critic and actor networks $Q'$ and $\mu'$ with parameters $\theta^{Q'}$ and $\theta^{\mu'}$ to give consistent targets to avoid divergence. Therefore, DDPG uses four neural networks. The critic and actor networks are updated sequentially in each iteration and their corresponding target networks are updated every $C$ steps. Instead of using the greedy policy to choose action at each iteration, \cite{lillicrap2015continuous} used the current actor policy and added noise sampled from a random process $\mathcal{N}$, so that $\bm{a_k}=\mu(\bm{s_k}|\theta_k^\mu)+\mathcal{N}$, making the action space to be continuous. $\theta_k^\mu$ is the parameter of the actor network $\mu$ at time $k$. To guarantee this estimation problem yields a stationary Markov Decision Process, the state $\bm{s_k}$ in RL needs to explicitly incorporate information regarding system states of both the estimator $(\ref{eq:e_linear})$ and nonlinear system  $(\ref{nonlinear_basic})$ instead of just distance error. Hence, like what we have done in \cite{sun2023emulation}, we define the state $\bm{s_k}=\{\bm{\tilde x_k},\bm{\hat x_k}\}$, where $\bm{\hat x_k}$ is the estimator state and $\bm{\tilde x_k}$ is the measurement of the nonlinear system state at time $k$. We design the reward function to be the $L_2$ norm of the error between $\bm{\tilde x_k}$ and $\bm{\hat x_k}$, i.e., $r_k=||\bm{\tilde x_k}-\bm{\hat x_k}||^2$. Meanwhile, we change the loss function in DDPG algorithm to one that is similar to what we have used in \cite{sun2023emulation}. Noting that we already have a total of $M$ trajectories, the DDPG can be applied either in a pseudo-online way by using the existing dataset or exploring online learning to collect new data points.
		
		\begin{algorithm}
			\caption	{Learning optimal input to estimate nonlinear systems with multiple trajectories}\label{alg:general_DDPG}
			\begin{algorithmic}
				\State Initialize replay memory $D$;
				\State Initialize critic network $Q$ with parameter $\theta^Q$ and actor network $\mu$ with parameter $\theta^\mu$;
				\State Initialize target-$Q'$ function with parameter $\theta^{Q'}=\theta^Q $ and target $\mu '$ with weighs $\theta^{\mu '}=\theta^\mu$;
				
				\For{\texttt{episode m = 1,\dots,M}}
				\State Initialize a random process $\mathcal{N}$;
				\State Start from the initial state $\bm{s_{0}}=\{\bm{\hat x_{m,0}},\bm{\tilde x_{m,0}}\}$, which contains the system states of the estimator and nonlinear system of trajectory with number $m$; 
				\For{\texttt{$k= 0,\dots, L$}}
				\State Choose input $\bm{a_k}=\mu(\bm{s_k}|\theta^\mu)+\mathcal{N}(k)$, get reward $r_k$ and new state $\bm{s_{k+1}}=\{\bm{\hat x_{m,k}},\bm{\tilde x_{m,k}}\}$;
				\State Store $(\bm{s_k}, \bm{a_k}, r_k,\bm{s_{k+1}})$ as a memory cue to $D$;
				\State Sample random minibatch of cues  $(\bm{s_j}, \bm{a_j}, r_j,\bm{s_{j+1}}),j\in[0,k]$ from $D$;
				\State Apply the loss function $Loss(\theta^Q;\theta^{Q'})=\max\{l_{DDPG}(\theta^{Q'};\theta^{Q}),l_{MSBE}(\theta^{Q})\}$ to train critic $Q$ network (the explicit form of the loss function is given in the proof of Lemma \ref{lemma});
				\State Update the actor network by $\nabla_{\theta^\mu} J \approx  \nabla_aQ(\bm{s,a}|\theta^Q)|_{\bm{s=s_k,a=\mu(s_k)}}\nabla_{\theta^\mu}\mu(\bm{s}|\theta^\mu )|_{\bm{s=s_k}} $;
				
				\State Update target network every C steps:
    $\theta^{\mu'}\leftarrow\theta^{\mu}$ and $\theta^{Q'}\leftarrow\theta^{Q}$.
				\EndFor
				\EndFor
			\end{algorithmic}
		\end{algorithm}	
		\begin{lemma}
  \label{lemma}
The loss function in Algorithm \ref{alg:general_DDPG} is non-increasing.
			\end{lemma}
\begin{proof}
Given the loss function $Loss(\theta^Q;\theta^{Q'})=\max\{l_{DDPG}(\theta^Q;\theta^{Q'}),l_{MSBE}(\theta^Q)\}$, where $l_{DDPG}(\theta^{Q})$ $=(r_k+\gamma Q_{\theta^{Q'}}(s_{k+1},\mu{'}_{\theta}(s_{k+1}))- Q_{\theta^{Q}}(s_k,a_k))^2$ and $l_{MSBE}(\theta^{Q})=(r_k+\gamma Q_{\theta^{Q}}(s_{k+1},\mu{'}_{\theta}(s_{k+1}))- Q_{\theta^{Q}}(s_k,a_k))^2$ denotes the mean squared Bellman error. Since $\theta^{Q'}_{i+1}=\arg\min\limits_{\theta^Q}Loss(\theta^Q;\theta^{Q'}_i)$, the non-increasing relation is
	\begin{equation}
 \nonumber
		\begin{split}
			\min \limits_{\theta^Q}Loss(\theta^Q;\theta^{Q'}_{i+1})&\le Loss(\theta^{Q'}_{i+1};\theta^{Q'}_{i+1})= l_{MSBE}(\theta^{Q'}_{i+1})\\
			&\le Loss(\theta^{Q'}_{i+1};\theta^{Q'}_{i})=\min \limits_{\theta^{Q}}Loss(\theta^{Q};\theta^{Q'}_{i})
			\end{split}
	\end{equation}
\end{proof}
After learning from the trajectories of (\ref{nonlinear_basic}) by using Algorithm \ref{alg:general_DDPG}, we obtain a critic network and an actor network that best approximate the optimal Q-function $Q^*$ and policy function $\mu^*$, which satisfy
\begin{subequations}	
\begin{align}
Q^*(\bm{s_k,a_k})&=r_k+\gamma\max\limits_{\bm{a'}} Q^*(\bm{s_{k+1},a'})\label{eq:Q_function} \\
\mu^*(\bm{s_k})&=\arg\max\limits_{\bm{a}} Q^*(\bm{s_{k},a}). \label{eq:policy_function}
\end{align}
\end{subequations}

In contrast to the DQN algorithm, which exclusively learns the Q-function as defined in (\ref{eq:Q_function}) and necessitates subsequent policy calculation, the DDPG algorithm offers direct computation of the policy function (\ref{eq:policy_function}). This distinction in approach significantly impacts the efficiency and directness of policy determination in reinforcement learning contexts. Therefore, the next time step estimation of ($\ref{eq:discrete}$) is given by
\begin{align}
\label{eq:prediction}
     \bm{ \hat x_{k+1}} = PU( A_r \bm{\tilde\Psi}(\bm{\hat x_k}) + \bm{\hat a}(\bm{s_k}) ),
\end{align}
where $\bm{\hat a}$ is generated by the trained actor network $\mu$. 

\section{Transfer Learning-based Optimal Policy}
Given the computational inefficiency inherent in accumulating data and training distinct neural networks for each nonlinear system under investigation, we instead propose generating a generalized estimator capable of achieving linear approximations across systems exhibiting diffeomorphic relationships, thereby greatly enhancing computational feasibility. The notion of transfer learning that we pursue next will allow treatment of larger classes of nonlinear systems to which Koopman theory has been shown to apply \cite{lan2013linearization}.

Consider a diffeomorphism function $\bm{h}$ that transfers (\ref{nonlinear_basic}) to a new nonlinear system
	\begin{equation}
		\label{eq:transfer}
		\begin{split}
			\dot{\bm{z}}&=\bm{g(z)},\\
		\end{split}
	\end{equation}
where $\bm{z=h(x)}$, with $\bm{z}$ and $\bm{x}$ having the same  dimension. Our goal is to use the relationship between these two systems and the optimal policy $\bm{\hat a}$ that has been obtained from Algorithm (\ref{alg:general_DDPG}) applied to (\ref{nonlinear_basic}) to predict the the discrete form $\bm{z_{k+1}=G(z_k)}$ trajectory of (\ref{eq:transfer}). 
Based on (\ref{eq:e_linear}), there is 
\begin{equation}
\label{eq:e_linear_2}
\bm{\tilde\Psi(\bm{ h^{-1}({\bm{\hat z_{k+1}}})  })}=A_r\bm{\tilde\Psi}(\bm{h^{-1}}({\bm{\hat z_k}}))+\bm{  a(\bm{h^{-1}({\bm{\hat z_k}}),h^{-1}(\bm{\tilde z_k})})}.
\end{equation}
Based on (\ref{eq:prediction}) and (\ref{eq:e_linear_2}), it suffices to predict the next state $\bm{\hat z_{k+1}}$ by
\begin{equation}
    	\label{eq:e_linear_3}
	\begin{split}
 \bm{\hat z_{k+1}}&= \bm{h}\left(  PU\left( A_r\bm{\tilde\Psi(\bm{h^{-1}({\bm{\hat z_k}}))}}+\bm{  \hat a_z}\right)  \right),\\
	\end{split}
\end{equation}
where $\bm{\hat a_z}$ is short for $ \bm{\hat a}(\bm{h^{-1}({\bm{\hat z_k}}),h^{-1}(\bm{\tilde z_k})})$.

\begin{theorem}
    Given $\bm{\hat a}$ such that $\|\bm{\hat a(s)}- \bm{a(s)}\|_2^2\le \epsilon$ for all $\bm{s}$, assuming $\bm{h}$ is $K$-Lipschitz with respect to $\|\cdot\|_2^2$, we have $\|\bm{z_{k+1}} - \bm{\hat z_{k+1}}\|_2^2\le K\epsilon$.
\end{theorem}
\begin{proof}
The optimal policy $\bm{a(s)}$ satisfies the equation $\bm{x_{k+1}}=PU(A_r\bm{\tilde\Psi(\hat x_k)}+\bm{a(s_k)})$ for all $k$ and its approximation, $\bm{\hat a(s)}$, satisfies the equation $\bm{\hat x_{k+1}}=PU(A_r\bm{\tilde\Psi(\hat x_k)})+\bm{\hat a(s_k)}$, where $P=[\bm{I}_n,\mathbf{0}_{N-n}]$ and $U^TU=\bm{I}_r$. Suppose $\|\bm{\hat a(s)}- \bm{a(s)}\|_2^2\le \epsilon$ for all $\bm{s}$, then $\|\bm{x_{k+1}}-\bm{\hat x_{k+1}}\|_2^2=\|PU(\bm{\hat a(s_k)}- \bm{a(s_k)})\|_2^2\le \|U(\bm{\hat a(s_k)}- \bm{a(s_k)})\|_2^2=\|\bm{\hat a(s_k)}- \bm{a(s_k)}\|_2^2\le \epsilon$. Given the diffeomorphism relation between $\bm{z}$ and $\bm{x}$, we have $\bm{z_{k+1}}=\bm{h(x_{k+1})}$ and $\bm{\hat z_{k+1}}=\bm{h(\hat x_{k+1})}$. If $\bm{h}$ is $K$-Lipschitz with respect to $\|\cdot\|_2^2$, $\|\bm{z_{k+1}} - \bm{\hat z_{k+1}}\|_2^2=\|\bm{h(x_{k+1})}-\bm{h(\hat x_{k+1})}\|_2^2\le K\|\bm{x_{k+1}}-\bm{\hat x_{k+1}}\|_2^2\le K\epsilon$.

\end{proof}

Of course, using (\ref{eq:e_linear_3}) as an estimator requires calculating the inverse function $\bm{h(\cdot)}$ in each computation of $\bm{\hat z_{k+1}}$, which is computationally inefficient especially when $\bm{h}$ is complicated. Consequently, we re-utilize the dataset $\mathcal{D}$. Given any noise-perturbed data point $\bm{\tilde x^j}$ in the data pair $(\bm{x^j},\bm{y^j})\in\mathcal{D}$, we can construct a corresponding data point $\bm{\tilde z^j}=\bm{h(\tilde x^j)}$. Concurrently, executing the trained actor network using the $M$ known state trajectories yields sequences of data $\{\bm{\hat x^j}\}_{j=1}^{M\times L}$ in one-to-one correspondence with $\{\bm{\tilde x^j}\}_{j=1}^{M\times L}$. $\bm{\hat z^j}$ is constructed by $\bm{h(\hat x^j)}$. Then, linear representations $O_1$ and $O_{2}$ are designed to be
\begin{equation}
    	\label{eq:transfer_l}
	\begin{split}
     O_1 = &\arg\min_{O}\sum_{j}\| \bm{\tilde\Psi(h^{-1}(\tilde z^j) }  ) -  O \bm{\tilde\Psi(\bm{\tilde z^j} }  )   \|^2\\
    =&\arg\min_{O}\sum_{j}\|\bm{\tilde\Psi}(\bm{\tilde x^j}   ) -  O \bm{\tilde\Psi(h(\tilde x^j)}   )   \|^2\\
 \end{split}
\end{equation}
\begin{equation}
    	\label{eq:transfer_l2}
	\begin{split}
     O_{2} = &\arg\min_{O}\sum_{j}\| \bm{\hat a (h^{-1} (\hat z^j), h^{-1} (\tilde z^j))} -  O  \bm{\hat a (\hat z^j, \tilde z^j)}   \|^2\\
     =& \arg\min_{O}\sum_{j}\| \bm{\hat a (\hat x^j,\tilde x^j)} -  O  \bm{\hat a ( h(\hat x^j),h(\tilde x^j))}   \|^2.
 \end{split}
\end{equation}

In this case, it suffices to take $ O_1 \bm{\tilde\Psi(z)}$ as an approximation of $ \bm{\tilde\Psi(h^{-1}(z))}$.
Combining with (\ref{eq:e_linear_2}), we finally get 
\begin{equation}
    	\label{eq:e_linear_4}
	\begin{split}
\bm{\tilde\Psi(\hat z_{k+1})}&= O_1^{\dagger}(A_r\bm{\tilde\Psi(\bm{h^{-1}({\bm{\hat z_k}}))}}+\bm{  \hat a_z}) \\
&=O_1^{\dagger}A_rO_1\bm{\tilde\Psi(\hat z_k)}+O_1^{\dagger}O_{2}\bm{\hat a}(\bm{\hat z_k,\tilde z_k}),\\
\bm{\bm{\hat z}_{k+1}}&=PU\bm{\tilde\Psi(\hat z_{k+1})}.
	\end{split}
\end{equation}
Therefore, we can directly obtain an estimator of the transferred system (\ref{eq:transfer}) using equations (\ref{eq:e_linear_4}) without requiring neural network training or numerically evaluating $\bm{h}^{-1}$ at each time-step. However, this transferred policy may have lost information when calculating matrices $O_1$ and $O_{2}$. To increase the estimation accuracy,  a strategy leverages the transferred model parameters as an advanced starting point for a limited fine-tuning process. Specifically, initializing the neural architecture with the transferred policy and further applying the DDPG algorithm allows refining the accuracy with a reduced computational burden. A numerical example is presented in the next section.

\section{Simulation and Analysis}
In this section, we apply the methods to a toy example and also the widely studied Van der Pol oscillator.

The toy nonlinear system we used is (\ref{ex1}). The training dataset has 4000 data points generated from $M=20$ trajectories whose initial points are uniformly distributed within a 10-unit radius disk and the sampling time is $\Delta t=0.1s$, while the testing data encompasses three length-20 trajectories from random initial points constrained to the disk.  The Koopman operator approximation utilizes polynomial bases up to order ten as the lifting dictionaries to span an observable subspace. Both the actor and critic architectures within the DDGP algorithm comprise fully-connected neural networks with three hidden layers employing ReLU activations. For comparative validation, equivalent network structures are constructed to directly regress actions from nonlinear system states without Koopman eigenvalue regularization. Actor and critic networks in the DDPG algorithm with Koopman operator approximation and comparison networks are trained for 20 episodes on the training dataset and their target networks update every 10 steps. 
\begin{figure}
	\begin{center}
            \includegraphics[scale=0.25]{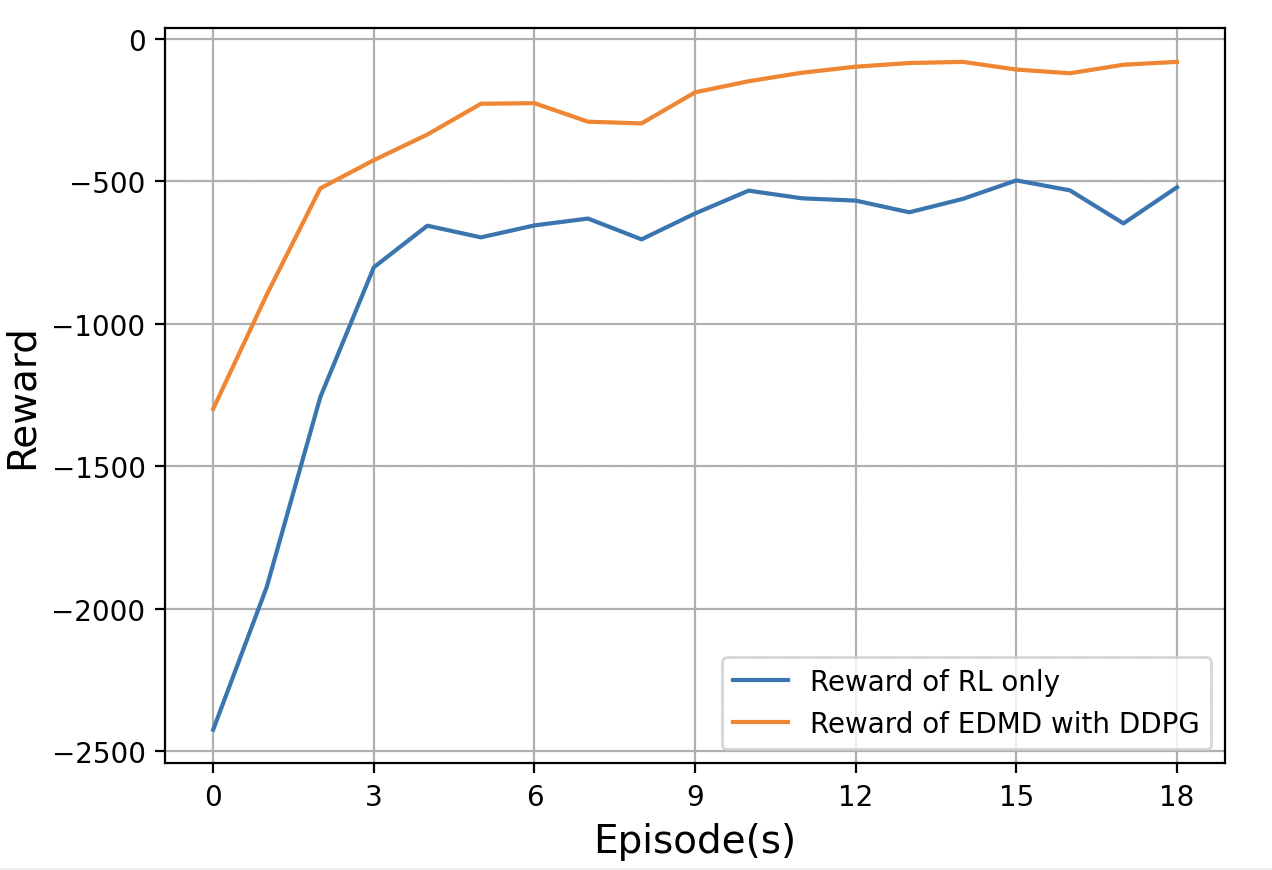}
		\includegraphics[scale=0.25]{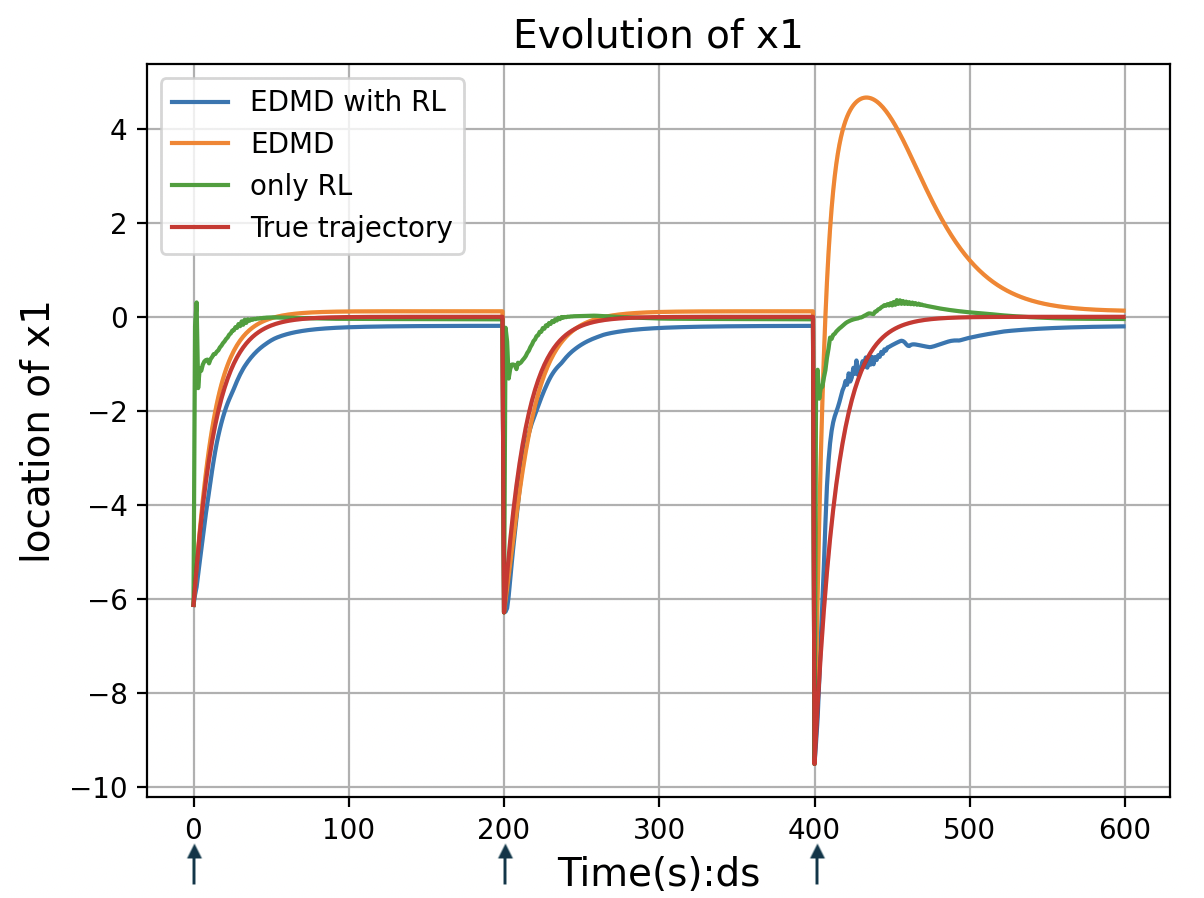}
            \includegraphics[scale=0.25]{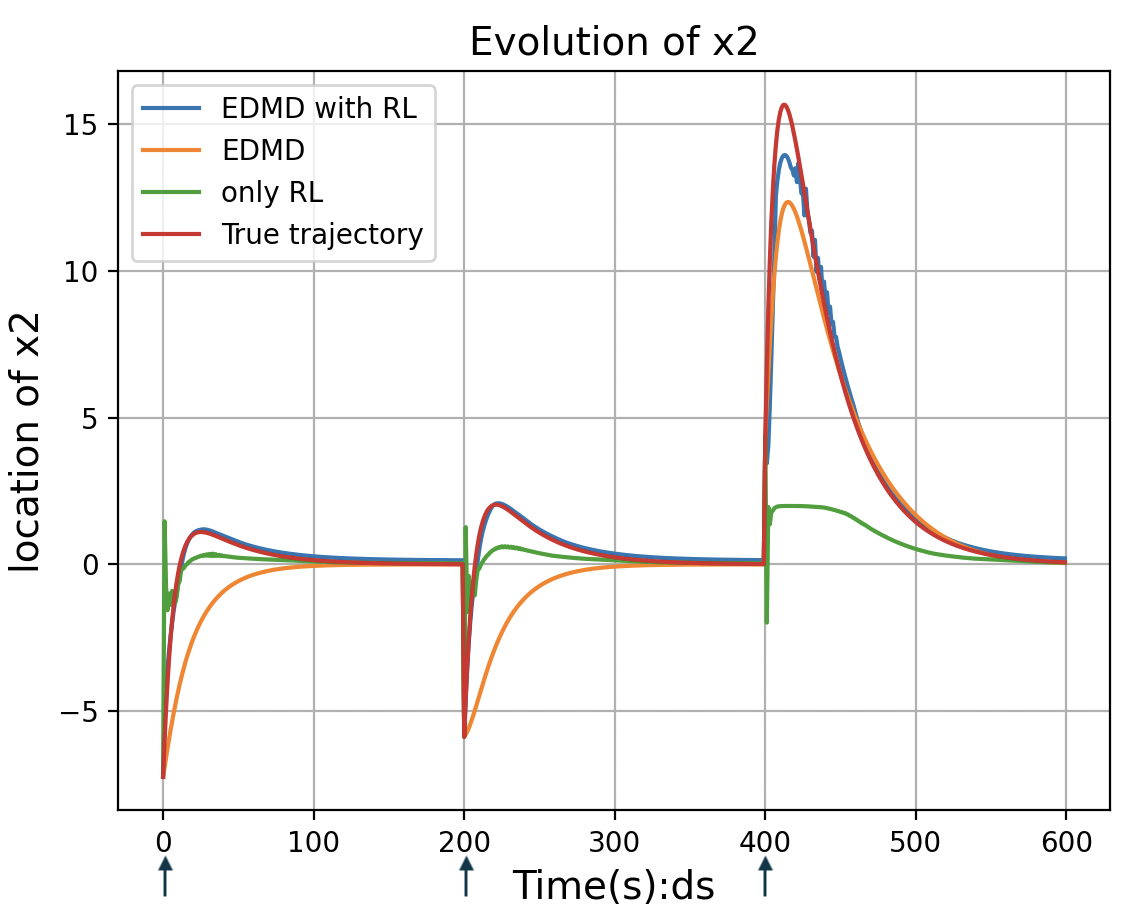}
	\end{center}
	\caption{The left figure exhibits the evolved reward during training using the EDMD merged with DDPG and only DDPG algorithm respectively. Figures on the right presented illustrate the performance of our estimator in approximating three distinct trajectories, each originating from different initial points. The arrows indicate that the estimator is initialized with the same starting points as the respective trajectories at the beginning.  }
	\label{fig:ex1}
\end{figure}

Quantitative results are presented graphically in the Fig. \ref{fig:ex1}. Each of the three trajectories in the testing set was evolved for 200 ds. It can be observed that incorporation of the EDMD within the reinforcement learning architecture demonstrably improves accumulated reward over the course of policy optimization. Given the underlying learning process, the reward evolution of the two methods (RL-only verses EDMD with RL) during training is predictable: a standalone reinforcement learning solution must exhaustively explore suboptimal action space before converging to optimal actions, whereas the our estimator only adjusts for residual dynamics uncharacterized during linear decomposition. Through partial modeling, learning is thus confined to a controlled correction minimizing discrepancies between the linear embedding's trajectory predictions and actual nonlinear flows. The framework thereby capitalizes on inherent system structure to expedite convergence. As shown in Fig. \ref{fig:ex1} (bottom), assessment on the testing dataset verifies superior generalization capability of the RL-baseed linear estimator versus alternatives. Meanwhile, the mean squared state estimation errors across different approaches are calculated, which is a quantitative metric used to evaluate the performance of an estimator by measuring the average squared deviation between the estimated state values ($x_1$ and $x_2$ in this example) and the true state values over the testing dataset. The proposed method achieved a mean squared state estimation error of 0.137, 
while EDMD and standalone reinforcement learning yielded significantly higher deviations of 2.289 and 3.096 respectively on the testing dataset. Notably, in instances where the linear embedding furnished by EDMD proves inadequate for fully characterizing the nonlinear dynamical flows, our proposed framework maintains comparatively precise reconstructions despite limitations in quantities of training data that might otherwise render the model underdetermined.

\begin{figure}[h]
	\begin{center}
            \includegraphics[scale=0.3]{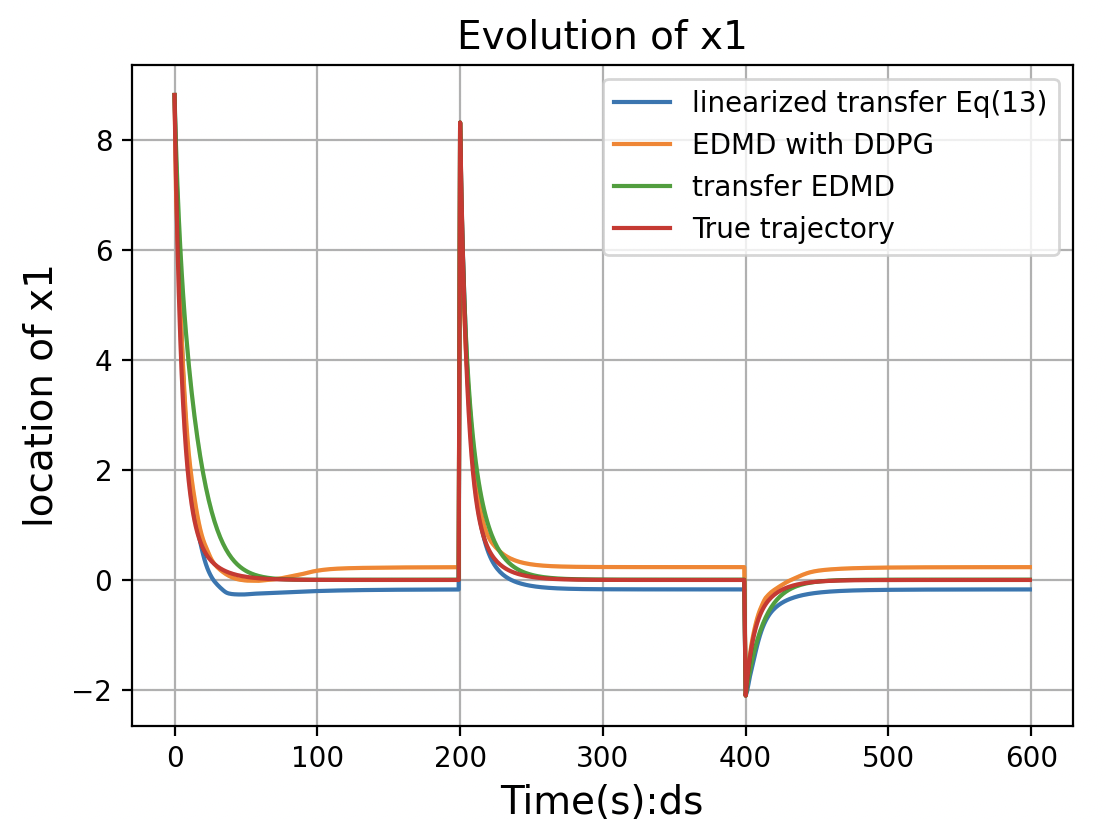}
            \includegraphics[scale=0.3]{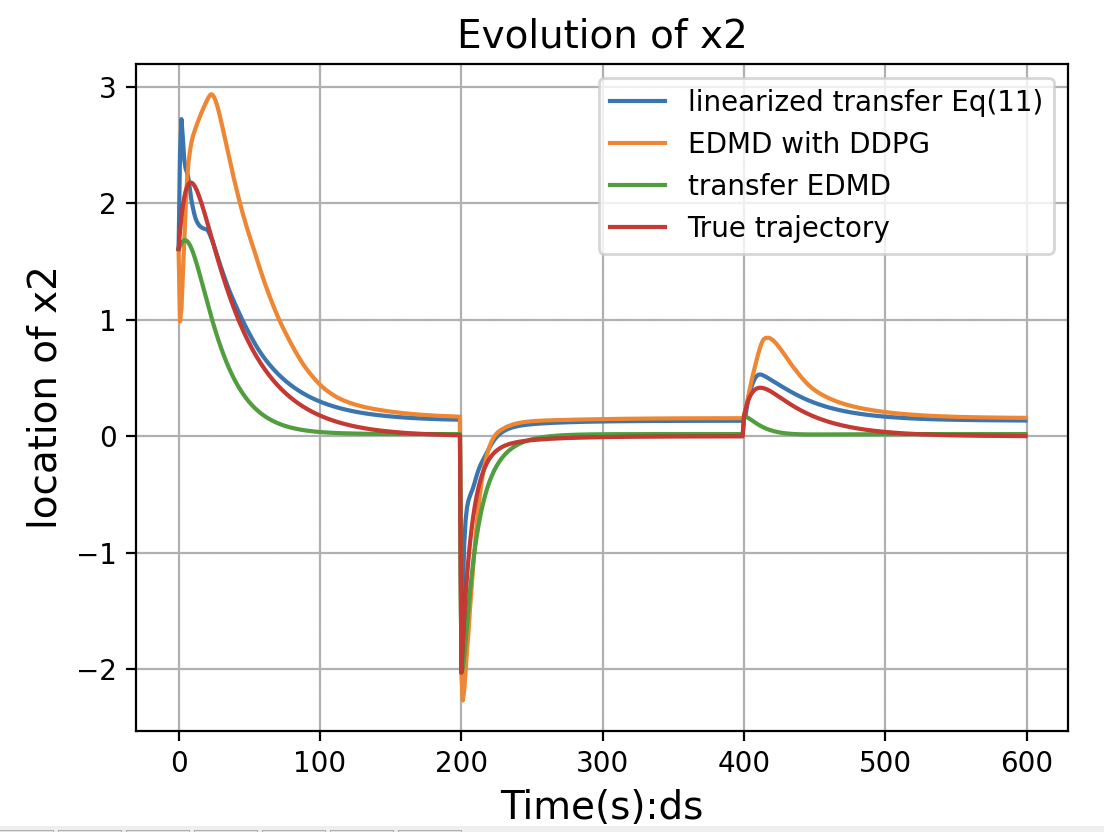}
	\end{center}
	\caption{The figure depicts the estimation performance by employing a transferred optimal policy in contrast to a newly trained estimator (\ref{eq:e_linear}) based on 20 episodes on a diffeomorphic nonlinear system. }
	\label{fig:transfer}
\end{figure}

    Suppose we have a diffeomorphism mapping $\bm{h}: \mathbb{R}^2 \rightarrow \mathbb{R}^2$ defined component-wise as $\bm{h}(x_1, x_2) = (x_1^3 + x_1, 2x_2)$, with an analytically available inverse transformation $\bm{h}^{-1}$ obtained by solving the corresponding cubic equation. Consider the lifted coordinate system $\bm{z} = \bm{h}(\bm{x})$. The dataset $\{\bm{z^j}\}$ and $\{\bm{\hat z^j}\}$ employed for analysis are derived from the original state trajectories $\{\bm{x^j}\}$ and $\{\bm{\hat{x}^j}\}$ corresponding to the system (\ref{ex1}). According to equations (\ref{eq:transfer_l}) and (\ref{eq:transfer_l2}), matrices $O_1$ and $O_{2}$ are easily calculated and the transferred estimator (\ref{eq:e_linear_4}) is obtained. To establish a baseline for comparative evaluation, we also consider an estimator (\ref{eq:e_linear}) trained on the transformed dataset $\{\bm{z^j}\}$ with 20 episodes of the DDPG algorithm. Results are shown in Fig. \ref{fig:transfer}. 

\begin{table*}
\begin{center}   
\caption{Average reward over the first 12 training episodes}  
\label{table:1} 
\begin{tabular}{|c|c|c|c|c|c|c|c|c|c|c|c|}   
\hline  \textbf{Episode(s)}  & 1 & 2& 3 & 4& 5 & 6& 7 & 8& 9 & 10& 11 \\   
\hline   \textbf{Random Initialization} & -462.3 &-367.3 & -223.5 & -170.5& -116.9 & -127.2& -97.6 & -52.2& -62.1 & -51.4& -54.1    \\ 
\hline   \textbf{Transferred Initialization} & -144.1 & -67.5& -67.7 & -66.4& -41.5 & -36.6& -36.9 & -38.0& -37.6 & -32.7& -33.2    \\      
\hline   
\end{tabular}   
\end{center}   
\end{table*}
It can be observed that the transferred estimator on the testing dataset reveals satisfactory generalization performance, attaining a mean squared error (MSE) of 0.0332. This represents a satisfactory performance over the scenario of training an estimator from scratch, which yields an MSE of 0.0952 on the same dataset. Moreover, we record execution times required for state estimation using the transferred policy versus the new trained estimator model. On an Apple M2 chip, the computational costs are 6.15 seconds versus 344 seconds respectively. The transferred estimator thus exhibits dual advantages of maintaining accuracy alongside significantly accelerated inference. While increasing the number of training episodes for the estimator (\ref{eq:e_linear}) can theoretically improve prediction fidelity, this would incur considerable overhead due to the repeated optimization procedure. An alternative strategy is applying transferred model parameters as initialization of the neural network. The quantitative results in Table \ref{table:1} demonstrate the merits of this transfer learning approach relative to handling each new scenario from scratch.

\begin{figure}[h]
	\begin{center}
            \includegraphics[scale=0.35]{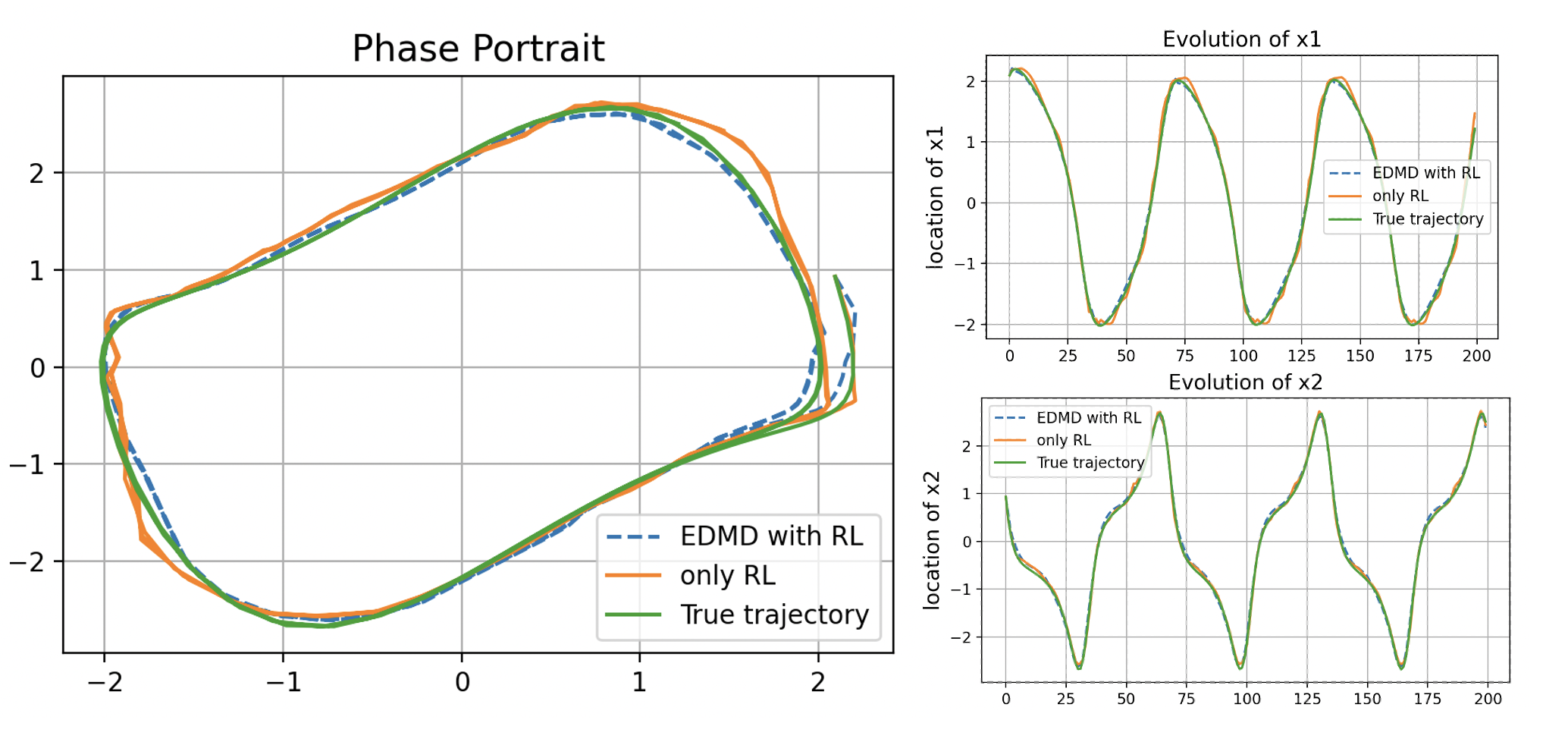}
	\end{center}
	\caption{The figure exhibits the performance of our approach on the Van der Pol equation and compares the stand-alone RL approach with hybrid combined EDMD plus RL. }
	\label{fig:ex2}
\end{figure}

To assess estimator performance on nonlinear oscillatory dynamics of a well-known system, we turn our attention to the Van der Pol oscillator, which is a second-order dynamic system with non-linear damping that has been widely used in the physical and biological sciences. The dynamics is described by $\ddot{y}-\alpha(1-y^2)\dot y +y=0$, whose state space form can be written as
\begin{equation}
\begin{split}
\dot x_1&=x_2\\
\dot x_2&=\alpha(1-x_1^2)x_2-x_1
\end{split}
\end{equation}
The training data comprises 20 trajectories generated from the Van der Pol system with physical parameter $\alpha=1$, and a 30dB white Gaussian noise has been added to the training data. Actually, the emulation performance does not change much with/without such a noise, but by adding noise may prevent over-fitting in some cases. Initial conditions are randomly sampled from a plausible domain, remaining unknown to the learning architectures. A separate test trajectory with random initialization for the Van der Pol system serves for evaluation, using a distinct random initialization. The result is presented in Fig. \ref{fig:ex2}, depicting the phase portrait and the evolution of states. 

For this representative example of a system containing a limit cycle, both the EDMD-regularized reinforcement learning approach and the RL stand-alone method converge to reasonable estimations after 10 training episodes. However, the proposed hybrid framework achieves superior accuracy, attaining mean squared errors of 0.0032 compared to 0.0048 for the reinforcement learning formulation operating directly on state-action pairs. Thus while neural networks can generate models that approximate the Van der Pol flows, incorporation of the Koopman eigengeometries provides more accurate trajectories.

\section{Conclusion}
This paper introduces a data-driven linear estimation framework for nonlinear dynamical systems based on the Koopman operator theoretic principles with reinforcement learning to compensate for residual unmodeled components. In fact, in the first step, EDMD serves merely as an illustrative approach and is not unique. It can be substituted by other Koopman methods, provided they are capable of extracting the dominant features from the data to construct Koopman operators.
Merged with reinforcement learning, our estimator provides an accurate yet interpretable representation of the nonlinear flow while generalizing across dynamics related by diffeomorphic transformations. Empirical evaluations on representative test cases, including the Van der Pol oscillator, quantify the estimator's predictive capabilities. Benchmarks validate significant reductions in mean squared error relative to purely linear embeddings or direct policy approximations, confirming the benefits of data-driven correction. While the current work focused on fully-observed deterministic autonomous systems, ongoing research aims to generalize the framework to systems with external inputs and partially observed with stochastic scenarios.  

\bibliographystyle{IEEEtran}  
\bibliography{root}
		
\end{document}